\documentclass{sig-alternate}
\usepackage[T1]{fontenc} 
\usepackage[utf8]{inputenc}
\usepackage[table]{xcolor}
 \usepackage{array,multirow,graphicx} 
\usepackage{tabularx}
\usepackage{balance}
\usepackage{pdflscape}
\usepackage{float}
\usepackage{balance}
\usepackage{xspace} 
\usepackage{hyperref}
\usepackage{listings}
\usepackage{framed}
\usepackage{amsmath}
\usepackage{amssymb}
\usepackage{datatool}
\usepackage{textpos}
\usepackage[linesnumbered]{algorithm2e}
\usepackage{numprint}
\usepackage{listings} 
\usepackage{enumitem}
\usepackage{color}
\usepackage[font=footnotesize,labelfont=bf]{caption} 
\pdfoutput=1

\newtheorem{definition}{Definition}

\begin{document}

\newcommand{\FIXME}[1]{{\color{red}#1}}
\newcommand{\fixme}[1]{{\color{red}#1}}
\newcommand{\nbprograms}{\numprint{7}\xspace}
\newcommand{\nbclasses}{\numprint{7}\xspace}
\newcommand{\nbsosies}{\numprint{472}\xspace}
\newcommand{\nbsosiesMR}{\numprint{12291s}\xspace}
\newcommand{\transfo}[1]{``#1''}
\newcommand{\checkablenumber}[1]{\FIXME{#1}}

\makeatletter
\def\@copyrightspace{\relax}
\makeatother

\newcommand\toolname{{DSpot}\xspace}
\title{\toolname: Test Amplification for Automatic Assessment of Computational Diversity}

\numberofauthors{1}
\author{
Benoit Baudry$^o$, Simon Allier$^o$, Marcelino Rodriguez-Cancio$^{\dag\dag,o}$ and Martin Monperrus$^{\dag,o}$\\
$^o$ Inria, France\\ 
$^{\dag\dag}$ University of Rennes 1, France
$^\dag$ University of Lille, France\\
contact: benoit.baudry@inria.fr}

\maketitle

\begin{abstract}
\textbf{Context:} Computational diversity, i.e., the presence of a set of programs that all perform compatible services but that exhibit behavioral differences under certain conditions, is essential for fault tolerance and security.
 
\textbf{Objective:} We aim at proposing an approach for automatically assessing the presence of computational diversity.
In this work, computationally diverse variants are defined as (i) sharing the same API, (ii) behaving the same according to an input-output based specification (a test-suite) and (iii) exhibiting observable differences when they run outside the specified input space. 

\textbf{Method:} Our technique relies on test amplification. We propose source code transformations on test cases to explore the input domain and systematically sense the observation domain. We quantify computational diversity as the dissimilarity between  observations on inputs that are outside the specified domain. 
 
\textbf{Results:} We run our experiments on \nbsosies variants of \nbclasses classes from open-source, large and thoroughly tested Java classes. Our test amplification multiplies by ten the number of input points in the test suite and is effective at detecting software diversity.

\textbf{Conclusion:} The key insights of this study are: the systematic exploration of the observable output space of a class provides new insights about its degree of encapsulation;  the behavioral diversity that we observe originates from areas of the code that are characterized by their  flexibility (caching, checking, formatting,  etc.).
\end{abstract}


\section{Introduction}
\label{sec:intro}

Computational diversity, i.e., the presence of a set of programs that all perform compatible services but that exhibit behavioral differences under certain conditions, is essential for fault tolerance and security \cite{avizienis85,deswarte98,donnell04,franz10}. Consequently, it is of utmost importance to have systematic and efficient procedures to determine if a set of programs are computationally diverse.

Many works have tried to tackle this challenge, 
using input generation \cite{jiang09}, static analysis \cite{kawaguchi2010conditional}, or evolutionary testing \cite{yoo2012} and \cite{Carzaniga15} (concurrent of this work). Yet, having a reliable detection of computational diversity for large object-oriented programs is still a challenging endeavor.

In this paper, we propose an approach, called \toolname\footnote{\toolname stands for \textbf{d}iversity \textbf{spot}ter}, for assessing the presence of computational diversity, i.e., to determine if a set of program variants exhibit different behaviors under certain conditions.
\toolname takes as input a test suite and a set of $n$ program variants. 
The $n$  variants have the same application programming interface (API) and they all pass the same test suite (i.e. they comply with the same executable specification). 
\toolname consists of two steps: (i) automatically transforming the test suite; and (ii) running this larger test suite, that we call ``amplified test suite'' on all variants to reveal visible differences in the computation.

The first step of \toolname is an original technique of test amplification \cite{xie06,zhang2012,pezze2013,yoo2012}. 
Our key insight is to combine the automatic exploration of the input domain with the systematic sensing of the observation domain.
The former is obtained by transforming the input values and method calls of the original test.
The latter is the result of the analysis and transformation of the original assertions of the test suite, in order to observe the program state from as many observation points visible from the public API as possible.
The second step  of \toolname runs the augmented test suite on each variant. The observation points introduced during amplification generate new traces on the program state. 
If there exists a difference between the trace of a pair of variants, we say that these variants are computationally diverse.
In other words, two variants are considered diverse if there exists at least one input outside the specified domain that triggers different behaviors on the variants which can be observed through the public API.  

To evaluate  the ability of \toolname at observing computational diversity, we consider \nbprograms open-source software applications. For each of them, we create \nbsosies  program variants, and we manually check that they are computationally diverse, they form our ground truth. 
We then run \toolname for each program variant.
Our experiments show that \toolname detects 100\% of the \nbsosies computational diverse program variants. 
In the literature, the technique that is the most similar to test amplification is by Yoo and Harman \cite{yoo2012}, called ``test data regeneration'' (TDR for short), we use it as baseline.
We show that test suites amplified with \toolname detect twice more  computationally diverse programs than TDR. In particular, we show that the new test input transformations that we propose bring a real added value with respect to TDR, to spot behavioral differences.

\medskip
To sum up, our contributions are:
\begin{itemize}[leftmargin=.4cm,itemsep=-.1cm]
  \item an original set of test cases transformations for the automatic amplification of an object-oriented test suite. 
  \item a validation of the ability of amplified test suites to spot computational diversity in  \nbsosies variants of \nbprograms open-source large scale programs.
  \item a comparative evaluation against the closest related work \cite{yoo2012}
  \item original insights about the natural diversity of computation due to randomness and variety of runtime environments.
  \item a publicly available implementation \footnote{\url{http://diversify-project.github.io/test-suite-amplification.html}} and benchmark \footnote{\url{http://diversify-project. eu/data/}}.
\end{itemize}

The paper is organized as follows: section \ref{sec:background} expands on the background and motivations for this work; section \ref{sec:approach} describes the core technical contribution of the paper: the automatic amplification of test suites; section \ref{sec:eval} presents our empirical findings about the amplification of \nbprograms real-world test suites and the assessment of diversity among \nbsosies program variants.

\section{Background}
\label{sec:background}

\begin{figure}
  \centering
  \includegraphics[width=0.85\columnwidth]{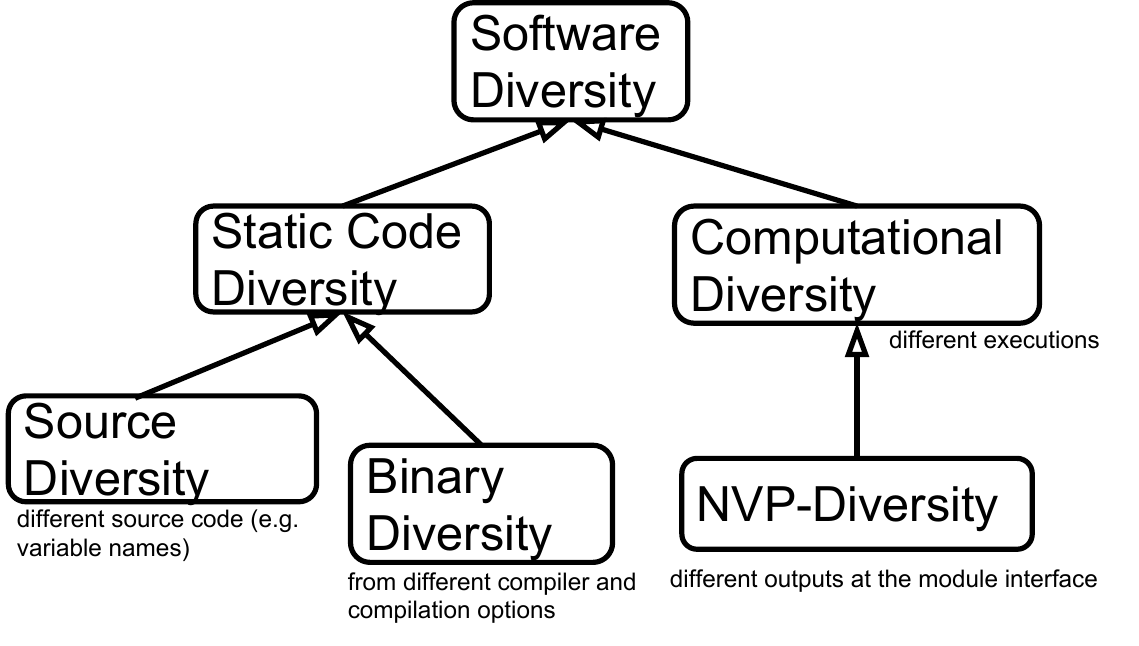}
  \caption{An High-level View of Software Diversity.}
  \label{fig:view-diversity}
\end{figure}

In this paper, we are interested in computational diversity.
Computational diversity is one kind of software diversity.
Figure \ref{fig:view-diversity} presents a high-level view of software diversity.
Software diversity can be observed statically either on source or binary code. 
Computational diversity is the one that happens at runtime. 
The computational diversity we target in this paper is NVP-Diversity, which relates to N-version programming.
It can be loosely defined as computational diversity that is visible at the module interface: different outputs for the same input.

\begin{lstlisting}[caption={Two subtraction functions. They are NVP-Diverse: there exists some inputs for which the output are different.},label=lst:wbbb,float,language=java,numbers=left]
public int subtract1(int a, int b) {
  return a-b;
}
public int subtract2(int a, int b) throws OverFlowException {
  BigInteger bigA = BigInteger.valueOf(a);
  BigInteger bigB = BigInteger.valueOf(b);
  BigInteger result = bigA.subtract(bigB);
  if (result.lowerThan(Integer.MIN_VALUE))) {
   throw new DoNotFitIn32BitException();
  }
  // the API requires an 32-bit integer value
  return result.intValue();}}
\end{lstlisting}

\subsection{N-version programming}
\label{sec:nversion}
In the Encyclopedia of Software Engineering, N-version programming is defined as  ``\emph{a software structuring technique designed to permit software to be fault-tolerant, ie, able to operate and provide correct outputs despite the presence of faults}'' \cite{knight1990n}.
In N-version systems, $N$ variants of the same module, written by different teams, are executed in parallel.
The faults are defined as an output of one or more variants that differ from the majority's output.
Let us consider a simple example with 2 programs, $p_1$ and $p_2$, if one observes a difference in the output for an input $x$ -- $p_1(x) \neq p_2(x)$ -- then a fault is detected.

Let us consider the example of Listing  \ref{lst:wbbb}.
It shows two implementations of subtraction, which have been developed by two different teams: a typical N-version setup.
\texttt{subtract1} simply uses the subtraction operator.
\texttt{subtract2} is more complex, it leverages BigInteger objects to handle potential overflows.

The specification given to the two teams states that the expected input domain is $[-2^{16},2^{16}] \times [-2^{16},2^{16}]$. To that extent, both implementations are correct and equivalent. 
These two implementations are run in parallel in production using a N-version architecture. 

If a production input is outside the specified input domain, e.g. \texttt{subtract1($2^{32}$+1, 2)}, the behavior of both implementations is different and the overflow fault is detected.

\subsection{NVP-Diversity}
\label{sec:definition-diversity}

In this paper, we use the term NVP-Diversity to refer to the concept of computational diversity in N-version programming:

\textbf{Definition: } Two programs are NVP-diverse if and only if there exists at least one input for which the output is different.

Note that according to this definition, if two programs are 
equivalent on all inputs, they are not NVP-diverse.

In this work, we consider programs in mainstream object-oriented programming languages (our prototype handles Java software). 
In OO programs, there is no such thing, as ``input'' and ``outputs''. This requires us to slightly modify our definition of NVP-Diversity.

Following \cite{harman2013comprehensive}, we replace ``input'' by ``stimuli'' and ``output'' by ``observation''.
A stimuli is a sequence of method calls and their parameters on an object under test.
An observation is a sequence of  calls to specific methods, to query the state of an object (typically getter methods).
The \textbf{input space} $\mathcal{I}$ of a class $P$ is the set of all possible stimuli for $P$.
The \textbf{observation space} $\mathcal{O}$ is the set of all sets of observations.

Now, we can clearly define NVP-diversity for OO-programs.

\textbf{Definition: } Two classes are NVP-diverse if and only if there exists two respective instances that produce different observations for the same stimuli.

\begin{figure}
  \centering
  \includegraphics[width=0.85\columnwidth]{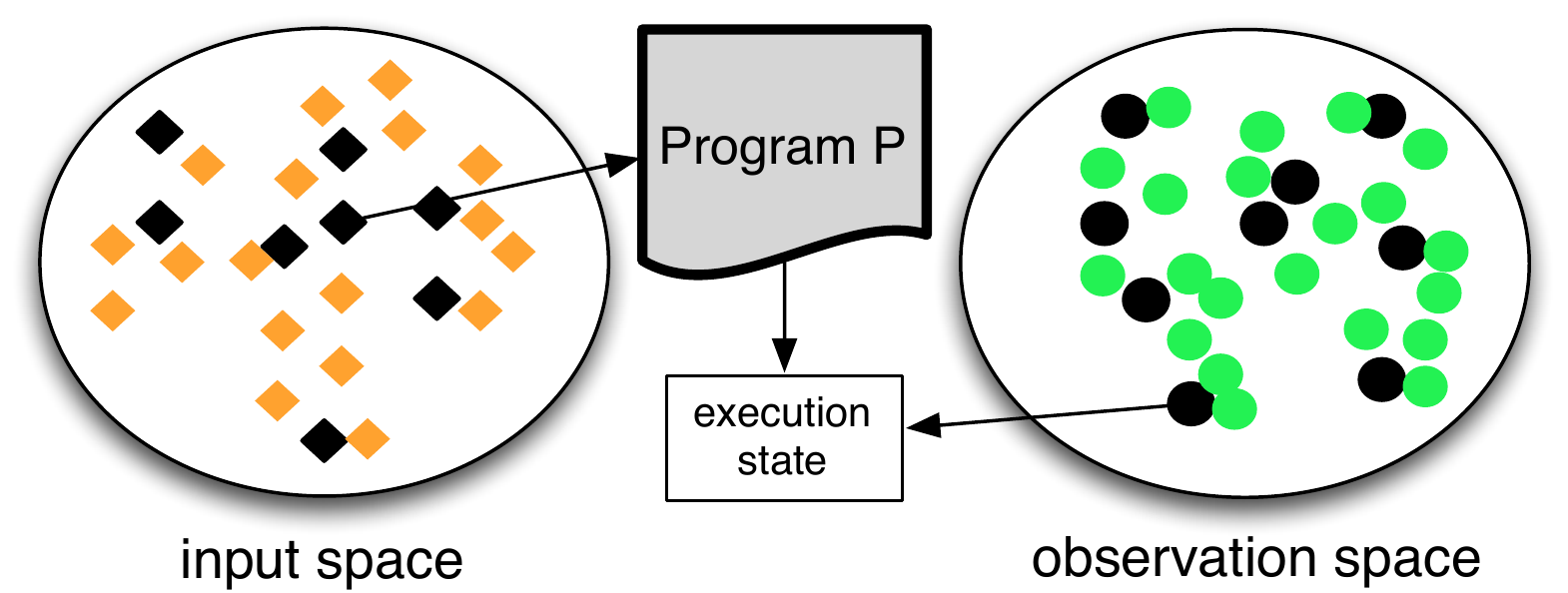}
  \caption{Original and amplified points on the input and observation spaces of P.}
  \label{fig:io-exploration}
\end{figure}

\subsection{Graphical Explanation}
\label{sec:unspecified}

The notion of NVP-diversity is directly related to activity of software testing as illustrated in figure \ref{fig:io-exploration}. 
The first part of a test case, incl. creation of objects and method calls, constitutes the stimuli, i.e. a  point in the program's input space (black diamonds in the figure). 
An oracle in the form of an assertion invokes one method and compares the result to an expected value: this constitutes an observation point on the program state that has been reached when running the program with a specific stimuli, the observation points of a test suite are black circles in the right hand side of the figure. 
To this extent, we say that a test suite specifies a set of relations between points in the input and observation spaces. 

\subsection{Unspecified Input Space}

In N-Version programming, by definition, the differences that are observed at runtime happen for unspecified inputs, which we call  \textbf{the unspecified domain} for short. 
In this paper, we consider that the points that are not exercised by a test suite form the unspecified domain. They are the orange diamonds in the left-hand side of the figure.

\section{Our Approach to Detect Computational Diversity}
\label{sec:approach}

We present \toolname, our approach to detect computational diversity. 
This approach is based on test suite amplification through automated transformations of test case code.

\subsection{Overview}
\label{sec:overview}

\begin{figure}
  \centering
  \includegraphics[width=0.8\columnwidth]{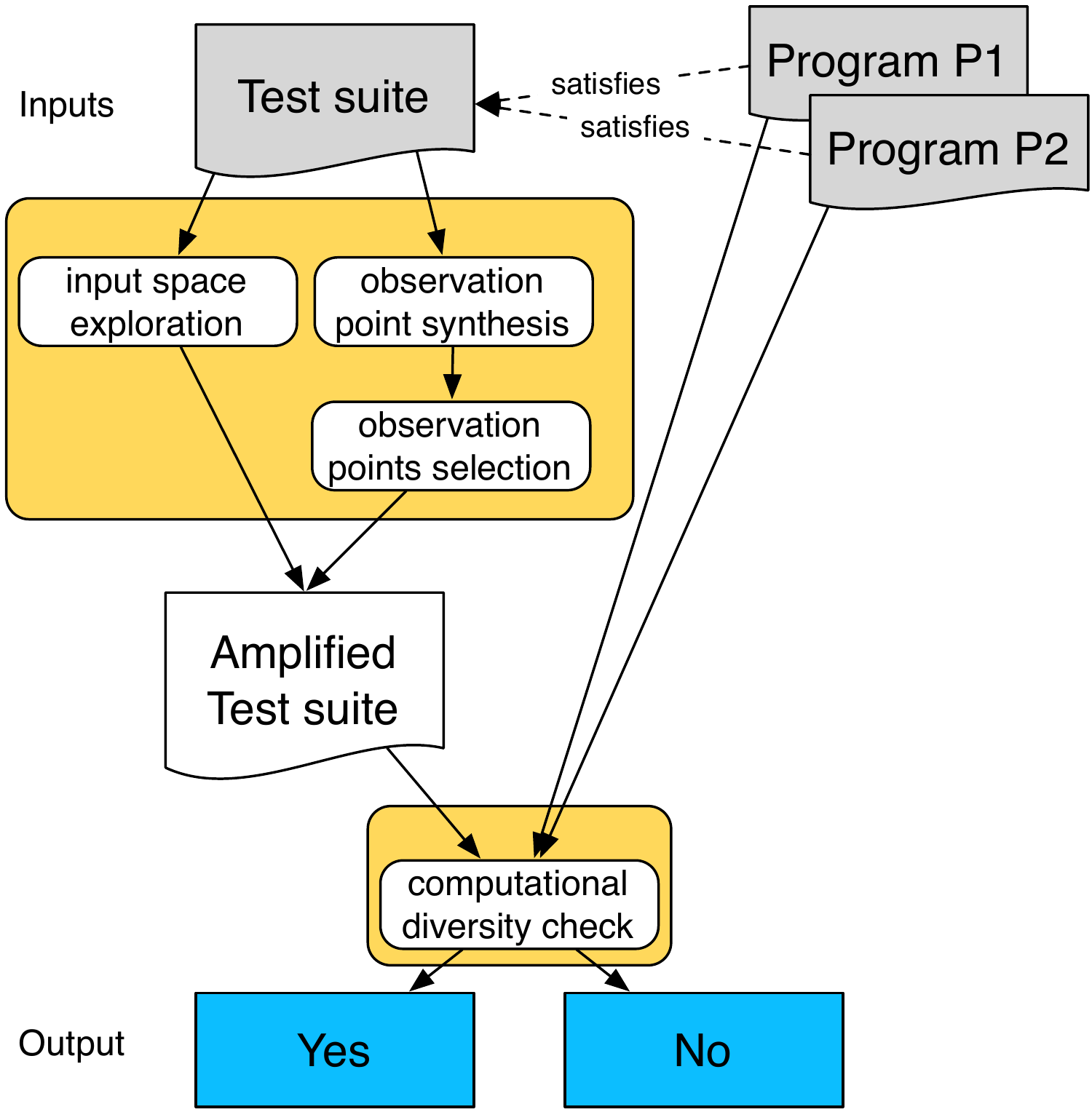}
  \caption{An overview of \toolname: a decision procedure for automatically assessing the presence of NVP-diversity.}
  \label{fig:overview}
\end{figure}

The global flow of \toolname is illustrated in figure \ref{fig:overview}. 

\textbf{Input:}
\toolname takes as inputs a set of program variants $P_1\ldots P_n$, which all pass the same test suite $TS$. 
Conceptually, $P_x$ can be written in any programming language.
There no assumption on the correctness or complexity of $P_x$, the only requirements is that they are all specified by the same test suite. In this paper,  we consider unit tests, however, the approach can be straightforwardly extended to other kinds of tests such as integration tests.

\textbf{Output:}
The output of \toolname is an answer to the question: are $P_1\ldots P_n$ NVP-diverse?

\textbf{Process:}
First, \toolname amplifies the test suite to explore the unspecified input and observation spaces (as defined in Section \ref{sec:background}). 
As illustrated in figure \ref{fig:io-exploration}, amplification generates new inputs and observations in the neighbourhood of the original points (new points are orange diamonds and green circles). This cartesian product of the amplified set of input and the complete set of observable points forms the amplified test suite $ATS$.

Also, Figure \ref{fig:overview} shows the step ``observation point selection'': this step removes the naturally random observations. 
Indeed, as discussed in more details further in the paper, some observations points produce diverse outputs between different runs of the same test case on the same program. This natural randomness comes from randomness in the computation and from specificities of the execution environment (addresses, file system, etc).

Once \toolname has generated an amplified test suite, it runs it on a pair of program variants to compare their visible behavior, as captured by the observation points. 
If some points reveal different values on each variant, they are considered as computationally diverse.

\subsection{Test Suite Transformations}

Our approach for amplifying test suites systematically explores  the neighbourhood of  the  input and observation points of the original test suite.
In this section we discuss the different transformations we perform for test suite amplification and algorithm \ref{alg:amplification} summarizes that procedure.
\begin{algorithm}[ht]
  \DontPrintSemicolon
  \KwData{$TS$ an initial test suite}
  \KwResult{$TS'$ an amplified version of TS}
  $TS_{tmp} \leftarrow \emptyset$\;
  \ForEach{$test\in TS$}
  {
 \label{alg:line:statements}
 \ForEach{$statement \in test$}
  {
  $test'\leftarrow$ clone(test)\;
  $TS_{tmp}\leftarrow$ remove(statement,test')\;
  $test''\leftarrow$ clone(test)\;
  $TS_{tmp}\leftarrow$ duplicate(statement,test'')\;
  }
  \ForEach{$literalValue \in test$}
  {
  $TS_{tmp}\leftarrow$ transform(literalValue,test)\; \label{alg:line:literals}
  }
  }
  $TS' \leftarrow {TS_{tmp} \cup TS}$
  \ForEach{$test \in TS'$}
  {
  removeAssertions(test)\;  \label{alg:line:asserts}
  }
  \ForEach{$test \in TS'$}
  {
  addObservationPoints(test)\; \label{alg:line:observations}
  }
  \ForEach{$test \in TS'$}
  {
  filterObservationPoints(test)\; \label{alg:line:filter}
  }
  \caption{Amplification of test cases}
  \label{alg:amplification}
\end{algorithm}

\subsubsection{Exploring the Input Space}

\textbf{Literals and statement manipulation:} The first step of amplification consists in transforming all test cases in the test suite with the following test case transformations.
Those transformations operate on literals and statements:
\begin{description}
  \item[Transforming literals:] given a test case $tc$, we run the following transformations for every literal value: a String value is transformed in three ways: remove, add a random character, and replace a random character by another one; a numerical value $i$ is transformed in four ways: $i+1$, $i-1$, $i\times2$, $i\div2$; a boolean value is replaced by the opposite value. These transformations are performed at line \ref{alg:line:literals} of algorithm \ref{alg:amplification}.
  \item[Transforming statement:] given a test case $tc$, for every statement $s$ in $tc$ we generate two test cases: one test case in which we remove s and another one in which we duplicate s. These transformations are performed at line \ref{alg:line:statements} of algorithm \ref{alg:amplification}.
\end{description}

Given the transformations described above, the transformation process has the following characteristics: (i)  each time we transform a variable in the original test suite, we generate a new test case (i.e., we do not `stack' the transformations on a single test case); (ii) the amplification process is exhaustive: given $s$ the number of String values, $n$ the number of numerical values, $b$ the number of booleans and $st$ the number of statements in an original test suite $TS$, \toolname produces an amplified test suite ATS of size: $|ATS|=s*3+n*4+b+st*2$. 

These transformations, especially the one on statements, can produce  test cases that cannot be executed (e.g., removing a call to \texttt{add} before a \texttt{remove} on a list). 
In our experiments, this accounted for approximately 10\% of the amplified test cases.

\textbf{Assertion removal:} The second step of amplification consists of removing all assertions from the test cases (line \ref{alg:line:statements} of algorithm \ref{alg:line:asserts}). 
The rationale is that the original assertions are here to verify the correctness, which is not the goal of the generated test cases. 
Their goal is to assess computational differences. 
Indeed, assertions that were specified for test case $ts$ in the original test suite are most probably meaningless for a test case that is variant of $ts$. 
When removing assertions, we are cautious to keep method calls that can be passed as a parameter of an \texttt{assert} method. 
We analyze the code of the whole test suite to find all assertions using the following heuristic: an assertion is a call to a method which name contains either \texttt{assert} or \texttt{fail} and which is provided by the JUnit framework. 
If one  parameter of the assertion  is a method call, we extract it, then we remove the assertion.
In the final amplified test suite, we keep the original test case, but also remove its assertion.

Listing \ref{new_TCs} illustrates the generation of two new test cases.
The first test method \texttt{testEntrySetRemoveChangesMap()} is the original one, slightly simplified for sake of presentation. 
The second one \texttt{testEntrySetRemoveChangesMap\_Add}, duplicates the statement \texttt{entrySet.remove} and does not contain the assertion anymore. 
The third test method \texttt{testEntrySetRemoveChangesMap\_DataMutator} replaces the numerical value 0 by 1.  

\begin{lstlisting}[caption={A test case \texttt{testEntrySetRemoveChangesMap} (\#1) that is amplified twice (\#2 and \#3)},label=new_TCs,language=java,numbers=left]
public void testEntrySetRemove() { // #1
  ...
  for (int i = 0; i < sampleKeys.length; i++) {
    entrySet.remove(new DefaultMapEntry<K, V>(sampleKeys[i], sampleValues[i]));
    assertFalse(
      "Entry should have been removed from the underlying map.",
      getMap().containsKey(sampleKeys[i]));
  } // end for
  ... }

public void testEntrySetRemove_Add() { // #2
  ...
  // call duplication
  entrySet.remove(new DefaultMapEntry<K, V>(sampleKeys[i], sampleValues[i]));
  entrySet.remove(new DefaultMapEntry<K, V>(sampleKeys[i], sampleValues[i]));
  getMap().containsKey(sampleKeys[i]);
  ... }

public void testEntrySetRemove_Data() { // #3
  ...
  // integer increment
  // int i = 0 -> int i = 1
  for (int i = 1 ; i < (sampleKeys.length) ; i++) {
    entrySet.remove(new DefaultMapEntry<K, V>(sampleKeys[i], sampleValues[i]));
    getMap().containsKey(sampleKeys[i]);
  } // end for
  ... }
\end{lstlisting}

\subsubsection{Adding Observation Points}

Our gaol is to observe different observable behaviors between a program and variants of this program.
Consequently, we need observation points on the program state. 
We do this by enhancing all the test cases in $ATS$ with observation points(line \ref{alg:line:observations} of algorithm \ref{alg:line:asserts}). These points are responsible for collecting pieces of information about the program state during or after the execution of the test case. 
In this context, an observation point is a call to a public method, which result is logged in an execution trace.

For each object $o$ in the original test case ($o$ can be part of an assertion or a local variable of the test case), we do the following:
\begin{itemize}
  \item we look for all getter methods in the class of $o$ (i.e., methods which name starts with \texttt{get}, that takes no parameter and whose return type is not void, and methods which name starts with \texttt{is} and return a boolean value) and call each of them. We also collect the values of all public fields.
  \item if the \texttt{toString} method is redefined for the class of $o$, we call it (we ignore the hashcode that can be returned by \texttt{toString})
  \item if the original assertion included a method call on $o$, we include this method call as an observation point.
\end{itemize}

\begin{table*}
  \caption{Descriptive Statistics about our Dataset}
\begin{tabularx}{\textwidth}{lp{5.5cm}p{3cm}XXXX}
  \hline
\textbf{Project}  & \textbf{Purpose} & \textbf{Class} & \textbf{LOC} & \textbf{\#tests} & \textbf{\scriptsize{coverage}} & \textbf{\#variants} \\
 commons-codec & Data encoding & Base64 & 255 &72 & 98\% & 12  \\
commons-collections  & Collection library  & TreeBidiMap&  1202 & 111 & 92\% &  133\\ 
commons-io &  Input/output helpers &  FileUtils& 1195 & 221 & 82\%&  44\\
commons-lang    & General purpose helpers (e.g. String) & StringUtils& 2247 & 233 & 99\%& 22 \\
guava   & Collection library &  HashBiMap & 525 & 35 & 91\%& 3  \\
gson   & Json library & Gson & 554 &  684& 89\%& 145 \\
JGit & Java implementation of GIT& CommitCommand & 433 & 138  & 81\% & 113\\
\hline
 \end{tabularx}
\label{tab:dataset}
\end{table*}

\textbf{Filtering observation points:} 
This introspective process provides a large number of observation points. 
Yet, we have noted in our pilot experiments that some of the values that we monitor change from one execution to another.
For instance, the identifier of the current thread  changes between two executions.
In Java, \texttt{Thread.currentThread().getId()} is an observation point that always  needs to be discarded for instance.

If we keep those naturally varying observation points,  \toolname would say that two variants are different while the observed difference would be due to randomness. This would be spurious results that are irrelevant for computational diversity assessment. 
Consequently, we discard certain observation points as follows.
We instrument the amplified tests $ATS$ with all observation points.
Then, we run $ATS$ 30 times on $P_x$, and repeat these 30 runs on three different machines. 
All observation points for which at least one value varies between at least two runs are filtered out (line \ref{alg:line:observations} of algorithm \ref{alg:line:filter}).

To sum up, \toolname produces an amplified test suite $ATS$ that contains more test cases than the original one in which we have injected observation points in all test cases.

\subsection{Detecting and Measuring the Visible Computational Diversity}

The final step of \toolname, runs the amplified test suite on pairs of program variants. 
Given $P_1$ and $P_2$,  the number of observation points  which  have a different values on each variant accounts for visible computational diversity. 
When we compare a set of variants, we use the mean number of differences over each pair of variants.

\subsection{Implementation}

Our prototype implementation amplifies Java source code~\footnote{the prototype is available here: \url{http://diversify-project.github.io/test-suite-amplification.html}}. The test suites are expected to be written using the JUnit testing framework, which is the \#1 testing framework for Java. It uses Spoon \cite{spoon} to manipulate the source code in order to create the amplified test cases. 
\toolname is able to amplify a test suite within minutes.

The main challenges for the implementation of \toolname were as follows: handle the many different situations that occur in real-world large test suites (use different versions of JUnit, modularize the code of the test suite itself, implement new types of assertions, etc.); handle large traces for comparison of computation (as we will see in the next section, we collect hundreds of thousands observations on each variant); spot the natural randomness in test case execution to prevent false positives in the assessment of computational diversity.

\section{Evaluation}
\label{sec:eval}

To evaluate whether \toolname is capable of detecting computational diversity, we set up a novel empirical protocol and apply it on \nbprograms large-scale Java programs.
Our guiding research question is:
\textbf{Is \toolname capable of identifying realistic large scale programs that are computationally diverse?}

\subsection{Protocol}
\label{sec:protocol}

First, we take large open-source Java programs that are equipped with good test suites.
Second, we forge variants of those programs using a technique from our previous work \cite{baudry14}. We call the variants sosie programs~\footnote{The word sosie is a French word that literally means ``look alike''}. 

\begin{definition}\label{def:sosie} \textbf{Sosie} (noun).
  Given a program $P$, a test suite $TS$ for $P$ and a program transformation $T$, a variant $P'$=$T(P)$ is a sosie of $P$ if the two following conditions hold
1) there is at least one test case in $TS$ that executes the part of $P$ that is modified by $T$
2) all test cases in $TS$ pass on $P'$.
\end{definition}

Given an initial program, we synthesize sosies with source code transformations that are based on the modification of the abstract syntax tree (AST).
As previous work \cite{legoues12,Schulte13}, we consider three
families of transformation that manipulate statement nodes of the AST:
1) remove a node in the AST (Delete);
2) adds a node just after another one (Add);
3) replaces a node by another one, e.g. a statement node is replaced by another statement (Replace). 
For ``Add'' and ``Replace'', the \textbf{transplantation point} refers to where a statement is inserted, the \textbf{transplant statement} refers to the statement that is copied and inserted and both transplantation and transplant points are in the same AST (we do not synthesize new code, nor take code from other programs). We consider transplant statements that manipulate variables of the same type as the transplantation point, and we bind the names of variables in the transplant to names that are in the namespace of the transplantation point. 
We call these transformations \emph{Steroid} transformations, and more details are available in our previous work \cite{baudry14}.

Once we have generated sosie programs, we manually select a set of sosies that indeed expose some computational diversity.
Third, we amplify the original test suites using our approach and also using a baseline technique by Yoo and Harman \cite{yoo2012} presented in \ref{sec:baseline}. 
Finally, we run  both amplified test suites and measure the proportion of variants (sosies) that are detected as computationally different.
We also collect additional metrics to further qualify the effectiveness of \toolname.

\subsection{Dataset}
We build a dataset of subject programs for performing our experiments. The inclusion criteria are the following:
1) the subject program must be real-world software;
2) the subject program must be written in Java;
3) the subject program's test suite must use the JUnit testing framework ;
4) the subject program must have a good test suite (a statement coverage higher than 80\%).

This results in Apache Commons Math, Apache Commons Lang, Apache Commons Collections, Apache Commons Codec and Google GSON and Guava. The dominance of Apache projects is due to the fact that they are among the very rare organizations with a very strong development discipline. 

In addition, we aim at running the whole experiments in less than one day (24 hours). Consequently we take a single class for each of those projects as well as all the test cases that exercise it at least once.

Table \ref{tab:dataset} provides the descriptive statistics of our dataset.
It gives the subject program identifier, its purpose, the class we consider, the class' number of lines of code (LOC), the number of tests that execute at least once one method of the class under consideration, the statement coverage and the total number of program variants we consider (excluding the original program).
We see that this benchmark covers different domains, such as data encoding and collections, and is only composed of well-tested classes. 
In total, there are between 12 and 145 computationally diverse variants of each program to be detected. This variation comes from the relative difficulty of manually forging computationally diverse variants depending on the project.

\begin{table*}[ht]
\centering
\caption{The performance of \toolname on amplifying \nbprograms Java test suites.}
\begin{tabularx}{\textwidth}{l|XX|XXX|X|p{0.7cm}}
\hline
  & \multicolumn{2}{c|}{Static}             &   \multicolumn{3}{c|}{Dynamic} & & \\
  & \#TC  & \#assert or obs.&  \#TC exec.  & \#assert or obs. exec. & \#disc. obs. & \# branch cov.    & \# path cov. \\
codec             & 72                & 509                 & 72    & 3528  &       & 124  & 	1245	\\
codec-DSpot       & 672 ($\times$9)   & 10597 ($\times$20)  & 672   & 16920 & 12    & 126   & 	12461 \\
collections       & 111               & 433                 & 768   & 7035  &       & 223   & 	376   \\
collections-DSpot & 1291 ($\times$12) & 14772 ($\times$34)  & 9202  & 973096& 0     & 224 & 	465   \\
io                & 221               & 1330                & 262   & 1346  &       & 366 & 	246   \\
io-DSpot          & 2518  ($\times$11)& 20408 ($\times$15)  & 2661  & 209911& 54313 & 373    & 	287   \\
lang              & 233               & 2206                & 233   & 2266  &       & 1014  & 	797   \\
lang-DSpot        & 988  ($\times$4)  & 12854  ($\times$6)  & 12854 & 57856 & 18    & 1015 &	  901   \\
guava             & 35                & 84                  & 14110 & 20190 &       & 60   & 	77    \\
guava-DSpot       & 625  ($\times$18) & 6834   ($\times$81) & 624656& 9464  & 0     & 60     & 	77    \\
gson              & 684               & 1125                & 671   & 1127  &       & 106  & 	84    \\
gson-DSpot        & 4992 ($\times$7)  & 26869  ($\times$24) & 4772  & 167150& 144   & 108 & 	137   \\
JGit              &138                & 176                 & 138   & 185   &       & 75   & 	1284  \\
JGit-DSpot        & 2152 ($\times$16) & 90828  ($\times$516)& 2089  & 92856 & 13377 & 75   & 	1735  \\
\hline
\end{tabularx}
\label{tab:test-amplification}
\end{table*}

\subsection{Baseline}
\label{sec:baseline}

In the area of test suite amplification, the work by Yoo and Harman \cite{yoo2012} is the most closely related to our approach. 
Their technique is designed for augmenting input space coverage but can be directly applied to detecting computational diversity.
Their algorithm, called test data regeneration -- TDR for short -- is based on four transformations on numerical values in test cases: 
data shifting ($\lambda x.x+1$ and  $\lambda x.x-1$ ) and data scaling (multiply or divide the value by 2) and a hill-climbing algorithm based on the number of fitness function evaluations.
They consider that a test case calls a single function, their implementation deals only with numerical functions and they consider the numerical output of that function as the only observation point.
In our experiment, we reimplemented the transformations on numerical values since the tool used by Yoo is not available. 
We remove the hill-climbing part since it is not relevant in our case. 
Analytically, the key differences between \toolname and TDR are:
TDR stacks mutliple transformations together;
\toolname has more new transformation operators on test cases:
 \toolname considers a richer observation space based on arbitrary data types and sequences of method calls.

\subsection{Research Questions}

We first examine the results of our test amplification procedure

\textbf{RQ1a: what is the number of generated test cases?}
We want to know whether our transformation operators on test cases enable us to create many different new test cases, i.e. new points in the input space. Since \toolname systematically explores all neighbors according to the transformation operators, we measure the number of generated test cases to answer this basic research question. 

\textbf{RQ1b: what is the number of additional observation points?}
In addition to creating new input points, \toolname creates new observation points. We want to know  the order of magnitude of the number  of those new observation points. 
To have a clear explanation, we start by performing only observation point amplification (without input point amplification) and count the total number of observations. We compare this number with the initial number of assertions, which exactly corresponds to the original observation points.

Then, we evaluate the ability of the amplified test suite to assess computational diversity. 

\textbf{RQ2a: does \toolname identify more computationally diverse programs than TDR?}
Now, we want to compare our technique with the related work. We count the number of variants that are identified as computationally different using \toolname and TDR. The one with with the highest value is better. 

\textbf{RQ2b: does the efficiency of \toolname come from the new inputs or the new observations?}
\toolname stacks two techniques: the amplification of the input space and the amplification of the observation space. To study their  impact  in isolation, we count the number of  computationally diverse program variants that are detected by the original input points equipped with new observation points and by the amplified set of input points with the original observations. 

The last research questions digs deeper in the analysis of amplified test cases and computationally diverse variants. 

\textbf{RQ3a: What is the number of natural randomness in computation?}
Recall that \toolname removes some observation points that naturally varies even on the same program. This phenomenon is due to the natural randomness of computation. To answer this question quantitatively, we count the number of discarded observation points, to answer it quantitatively, we discuss one case study. 

\textbf{RQ3b: what is the richness of computational diversity?}
Now, we really understand the reasons behind the computational diversity we observe. We take a random sample of three pairs of computationally diverse program variants and analyze them. We discuss our findings.

\subsection{Empirical Results}

We now discuss the empirical results obtained on applying \toolname on our dataset.

\begin{table*}[ht]
\centering
\caption{The effectiveness of computational diversity detection}
\begin{tabularx}{\textwidth}{lXXXXXX}
\hline
   & \#variants detected by \toolname & \#variants detected by TDR & input space effect & observation space effect & mean \# of divergences \\
commons-codec           & 12/12   & 10/12   &12/12    & 10/12   & 21.9 \\
commons-collections     & 133/133 & 133/133 & 133/133 & 133/133 &  5207.9  \\ 
commons-io              & 44/44   & 18/44   & 42/44   & 18/44   & 405.5 \\
commons-lang            & 22/22   & 0/22    & 10/22   & 0/22    &  22.9 \\
guava                   & 3/3     & 0/3     & 0/3     & 3/3     & 2  \\
gson                    &145/145  &0/145    & 134/145 & 0/145   & 801.5 \\
jgit                    &113/113  &0/113    & 113/113 & 0/113   & 1565.4 \\
\hline
\end{tabularx}
\label{tab:diversity-measures}
\end{table*}

\subsubsection{\# of Generated Test Cases}

Table \ref{tab:test-amplification} presents the key statistics of the amplification process. 
The lines of these table go by pair: one that provides data for one subject program and the following one that provides the same data gathered with the  test suite amplified by \toolname.
Columns from  2 to 5 are organized in two groups: the first group gives a static view on the test suites (e.g. how many test methods are declared); 
the second group draws a dynamic picture of the test suites under study (e.g. how many assertions are executed).

Indeed, in real, large-scale programs, test cases are modular. Some test cases are used multiple times because they are called by other test cases. For instance, a test case that specifies a contract on  a collection is called  when testing  all implementations of collections (ArrayList, LinkedList, etc.). We call them \emph{generic tests}.

Let's first concentrate on the static values.
Column 2 gives the number of  test cases in the original and amplified test suites, while column 3 gives the number of assertions in the original test suites and the number of observations in the amplified. 

One can see that our amplification process is massive. We create between 4x and 12x more test cases than the original test suites. 
For instance, the  test suite considered for commons.codec contains 72 test cases. \toolname produces an amplified test suite that contains 672 test methods: 9x more than the original test suite. The original test suite observes the state of the program with 509 assertions, while \toolname employs 10597 observations points to detect computational differences.

Let us now consider the dynamic part of the table. Column 4 gives the number of  tests executed (\#TC exec.) and column 5 the number of assertions executed or the number of observation points executed. 
Column 6 gives the number of the discarded observation points because of natural variations (discussed in more details in section \ref{sec:natural:randomness}).
As we can see, the number of generated tests (\#ATC exec.) is impacted by amplification.
For instance, for commons.collection there are 1291 tests in the amplified test suite, but altogether,  9202 test cases are executed.
The reason is that we synthesize new test cases that use other generic test methods. 
Consequently, this increases the number of executed generic test methods, which  is included in our count. 

Our test case transformations yield a rich exploration of the input space. 
Columns 7 to 11 of Table \ref{tab:test-amplification} provide deeper insigths about the synthesized test cases.
Colum 7 gives the branch coverage of the original test suites and the amplified ones (lines with *-DSPOT identifiers).
While  original test suites have a very high branch coverage rate, yet, DSpot is still able to generate new teststhat cover a few previously uncovered branches. 
For instance, the amplified test suite for commons-io/FileUtils reaches 7 branches that were not executed by the original test suite. 
Meanwhile, the original test suite for guava/HashBiMap already covers 90\% of the branches and DSpot did not generate test cases that cover new branches. 

The richness of the amplified test suite is also revealed in the last column of the table (path coverage): it provides the cumulative number of different paths executed by the test suite in all methods under test. 
The amplified test suites cover much more paths than the original ones, which means that they trigger a much wider set of executions of the class under test than the original test suites. 
For instance, for Guava, the total number of different paths covered in the methods under test increases from 84 to 137. 
This means that, while the amplified test suite does not cover many new branches, it executes the parts that were already covered in many novel ways, increasing the diversity of executions that are tested. 
There is one extreme case in  the \texttt{encode} method of commons-codec\footnote{line 331 in the Base64 class \url{https://github.com/apache/commons-codec/blob/ca8968be63712c1dcce006a6d6ee9ddcef0e0a51/src/main/java/org/apache/commons/codec/binary/Base64.java}}: the original test suite  covers 780 different paths in this method, while the amplified test suite  covers 11356 different paths. 
This phenomenon is due to the complex control flow of the method and to the fact that its behavior directly depends on the value of an array of bytes that takes many new values in the amplified test suite. 

\begin{framed}
The amplification process is massive and produces rich new input points: the number of declared and executed test cases and the diversity of executions from test cases increase. 
\end{framed}

\subsubsection{\# of Generated Observation Points}

Now we focus on the observation points.
The fourth column of Table \ref{tab:test-amplification} gives the number of assertions in original test suite. This corresponds to the number of locations where the tester specifies expected values about  the state of the program execution. 
The fifth column, gives the number of observation points in the amplified test suite. 
We do not call them assertions since they do not contain an expected value, i.e., there is no oracle. 
Recall that we use those observation points to compare the behavior of two program variants in order to assess the computational diversity. 

As we can see, we observe the program state on many more observation points than the original assertions. As discussed in Section \ref{sec:definition-diversity}, those observations points use the API of the program under consideration, hence allow to reveal visible and exploitable computational diversity.
However, this number also encompasses the observation points on the new generated test cases.

If we look at the dynamic perspective (second part of Table \ref{tab:test-amplification}), one observes the same phenomenon as for test cases and assertions, there are many more  points actually observed during test execution than statically declared ones. The reasons are identical, many observations points are in generic test methods that are executed several times, or are within loops in test code. 

\begin{framed}
These results validate our initial intuition that a test suite only covers a small portion of the observation space. It is possible to observe the program state from many other observation points.
\end{framed}

\subsubsection{Effectiveness}

We want to assess whether our method is effective for identifying computationally diverse program variants. 
As golden truth, we have the forged variants for which we know that they are NVP-diverse (see Section \ref{sec:protocol}), their numbers are given in the descriptive Table \ref{tab:dataset}. The benchmark is publicly available at \url{http://diversify-project.eu/data/}. 

We run \toolname and TDR to see whether those two techniques are able to detect the computationally diverse programs. 
Table \ref{tab:diversity-measures} gives the results of this evaluation.
The first column contains the name of the subject program. 
The second column gives the number of variants detected by \toolname.
The third column gives the number of variants detected by TDR.
The last three columns explore more in depth whether computational diversity is reveales by new input points or new observation points or both, we will come back to them later. 

As we can see, \toolname is capable of detecting all computationally diverse variants of our benchmark.
On the contrary, the baseline technique, TDR, is always worse. Either it detects only a fraction of them (e.g. 10/12 for commons.codec) or even not at all. 
The reason is that TDR, as originally proposed  by Yoo and Harman, focuses on simple programs with shallow input spaces (one single method with integer arguments). On the contrary, \toolname is designed to handle rich input spaces, incl. constructor calls, method invocations and strings. This has a direct impact on the effectiveness of detecting computational diversity in program variants.

Our technique is based on two insights: the amplification of the input space and the amplification of the observation space. 
We now want to understand the impact of each of them. 
To do so, we disable one or the other kind of amplification and measure the number of detected variants. 
The result of this experiment is given in the last two columns of 
Table \ref{tab:diversity-measures}. 
Column ``input space effect'' gives the number of variants that are detected only by the exploration of the input space (i.e. by observing the program state only with the observation method used in the original assertions).
Column ``observation space effect'' gives the number of variants that are detected only by the exploration of the observation space (i.e. by observing the result of method calls on the objects involved in the test).
For instance, for commons-codec, all variants (12/12) are detected by exploring the input space, and 10/12 are detected by exploring the observation space. This means that 10 of them are detected are detected either by one exploration or the other one. 
On the contrary for guava, only the exploration of the observation space enables \toolname to detect the three computationally diverse variants of our benchmark. 

By comparing columns ``input space effect'' and ``observation space effect'' one sees that our two explorations are not mutually exclusive and are complementary. Some variants are detected by both kinds of exploration (as in the case of commons-codec).
For some subjects, only the exploration of the input space is effective (e.g. commons-lang), while for others (guava), this is the opposite.
Globally, the exploration of the input space is more efficient, most variants are detected this way. 

Let us now consider the last column of Table \ref{tab:diversity-measures}. It gives the mean number of observation points for which we observe a difference between the original program and the variant to be detected. 
For instance, among the 12 variants for commons.codec, there is on average 21.9 observation points for which there is a difference. 
Those numbers are high,  showing that the observation points are not independent. Many of the methods we call to observe the program state inspect a different facet of the same state. For instance, in a list, the methods \texttt{isEmpty()} and \texttt{size} are semantically correlated.

\begin{framed}
The systematic exploration of the input and the observation spaces is effective at detecting behavioral diversity between program variants.
\end{framed}

\subsubsection{Natural Randomness of Computation}
\label{sec:natural:randomness}

When experimenting with \toolname on real programs, we noticed that some observation points naturally vary even  when running the same test case several times on the same program. 
For instance, a hashcode that takes into account a random salt can be different between two runs of the same test case. 
We call this effect, the ``natural randomness'' of test case execution.

We distinguish two kinds of natural variations in the execution of test suites. 
First, some observation points vary \emph{over time} when the test case is executed several times on the same environment (same machine, OS, etc.). This is the case for the hashcode example. 
Second, some observation points vary \emph{depending on the execution environment}. 
For instance, if one adds an observation point on a file name, the path name convention is different on Unix and Windows systems. If method \texttt{getAbsolutePath} is an observation point, it may return \texttt{"/tmp/foo.txt"} on Unix and \texttt{"C:\textbackslash tmp\textbackslash foo.txt"} on Windows. While this first example is pure randomness, the second only refers to variations in the runtime environment.

\begin{lstlisting}[caption={An amplified test case with observation points that naturally vary, hence are discarded by \toolname},label=monitor_ex,float,language=java,numbers=left]
void testCanonicalEmptyCollectionExists() {
  if (((supportsEmptyCollections()) && (isTestSerialization())) && (!(skipSerializedCanonicalTests()))) {
    Object object = makeObject();
    if (object instanceof Serializable) {
      String name = getCanonicalEmptyCollectionName(object);                  
      File f =     new java.io.File(name);
      // observation on f 
      Logger.logAssertArgument(f.getCanonicalPath());
      Logger.logAssertArgument(f.getAbsolutePath());
      .....
    }} 
}
\end{lstlisting}

Interestingly, this natural randomness is not problematic in the case of the original test suites, because it remains below the level of observation of the oracles (the test suite assertions in JUnit test suites).  However, in our case, if one keeps an observation point that is impacted by some natural randomness, this would produce a \textbf{false positive }for computational diversity detection. Hence, as explained in Section \ref{sec:approach},  one phase of \toolname consists in detecting the natural randomness first and discarding the impacting observation points. 

Our experimental protocol enables us to quantify the number of discarded observation points. The 6th column of Table \ref{tab:test-amplification} gives this number. For instance, for commons-codec, \toolname detects 12 observation points that naturally vary. 
This column shows two interesting facts. 
First, there is a large variation in the number of discarded observation points, it goes up to 54313 for commons-io. This case, together with JGIT (the last line), is due to the heavy dependency of the library on the underlying file system (commons-io is about IO -- hence file systems --operations, JGIT is about manipulating GIT versioning repositories that are also stored on the local file system).

Second, there are two subject programs (commons-collections and guava) for which we discard no points at all. In those programs, \toolname does not detect a single point that naturally varies by running 100 times the test suite on three different operating systems. 	The reasons is that the API of those subject programs does not allow to inspect the internals of the program state up to the naturally varying parts (e.g. the memory addresses). We consider this good as this, it shows that the encapsulation is good: more than providing an intuitive API, more than providing a protection against future changes, \emph{it also completely encapsulates the  natural randomness of the computation}.

Let us now consider a case study.
Listing \ref{monitor_ex} shows an example of an amplified test with  observation points for Apache Commons Collection. 
There are 12 observation methods that can be called on the object \texttt{f} instance of \texttt{File} (11 getter methods and \texttt{toString}). The figure shows two getter methods that return different values from one run to another (there are 5 getter methods with that kind of behavior for a \texttt{File} object). We ignore these observation points when comparing the original program with the variants.

\begin{framed}
The systematic exploration of the observable output space provides new insights about the degree of encapsulation of a class. 
When a class gives public access to variables that naturally vary, there is a risk that when used in oracles, they result in flaky test cases.
\end{framed}

\subsubsection{Nature of Computational Diversity}
\label{sec:nature-computational-diversity}

Now we want to understand more in depth the nature of the NVP-diversity we are observing.
Let us discuss three case studies.

\begin{lstlisting}[caption={Two variants of \texttt{writeStringToFile} in commons.io},label=lst:roci3,float,language=java,numbers=left]
//original program
void writeStringToFile(File file, String data, Charset encoding, boolean append) throws IOException {
  OutputStream out = null;
  out = openOutputStream(file, append); 
  IOUtils.write(data, out, encoding);
  out.close(); }

// variant  
void writeStringToFile(File file, String data, Charset encoding, boolean append) throws IOException {
  OutputStream out = null;
  out = new FileOutputStream(file, append);
  IOUtils.write(data, out, encoding);
  out.close(); }
\end{lstlisting}

\begin{lstlisting}[caption={Amplified test case that reveals computational diversity between variants of listing \ref{lst:roci3}},label=lst:roci3test,float,language=java,numbers=left]
void testCopyDirectoryPreserveDates() {
 try {
  File sourceFile = new File(sourceDirectory, "hello/txt");
  FileUtils.writeStringToFile(sourceFile, "HELLO WORLD", "UTF8");
 catch (Exception e) {
  DSpot.observe(e.getMessage());
  }
}
\end{lstlisting}

Listing \ref{lst:roci3} shows two variants of the \texttt{writeStringToFile()} method of Apache Commons IO.
The original program calls \texttt{openOutputStream}, which checks different things about the file name, while the variant  directly calls the constructor of \texttt{FileOutputStream}. These two variants  behave differently outside the specified domain: in case \texttt{writeStringToFile()} is called with an invalid file name, the original program handles it, while the variant throws a \texttt{FileNotFoundException}.
Our test transformation operator on String values produces such a file name, as shown in the test case of listing \ref{lst:roci3test}: a ``.'' is changed into a star ``/''. This made the file name an  invalid one. Running this test on the variant results in a \texttt{FileNotFoundException}.

\begin{lstlisting}[caption={Two variants of \texttt{toJson} in GSON},label=lst:roci2,float,language=java,numbers=left]
// Original program
void toJson(Object src, Type typeOfSrc, JsonWriter writer){
    writer.setSerializeNulls(oldSerializeNulls); } }
//variant
void toJson(Object src, Type typeOfSrc, JsonWriter writer){
    writer.setIndent("  ") 
} }
\end{lstlisting}

\begin{lstlisting}[caption={Amplified test detecting black-box diversity among variants of listing \ref{lst:roci2} },label=lst:roci2test,float,language=java,numbers=left]
 public void testWriteMixedStreamed_remove534() throws IOException {
   %* {\color{gray} \emph{...} } *)    
   gson.toJson(RED_MIATA, Car.class, jsonWriter);
   jsonWriter.endArray();
   Logger.logAssertArgument(com.google.gson.MixedStreamTest.CARS_JSON);
   Logger.logAssertArgument(stringWriter.toString());
   %* {\color{gray} \emph{...} } *)    
 }
\end{lstlisting}

Let us now consider listing \ref{lst:roci2}, which shows two variants of  the \texttt{toJson()} method from the Google Gson library. 
The last statement of the original method is replaced by another one: instead of setting the serialization format of the \texttt{writer} it set the indent format. 
Each variant creates a JSon with slightly different formats,  and none of these formatting decisions are part of the specified domain (and actually, specifying the exact formatting of the JSon String could be considered as over-specification). 
The diversity among variants is detected by the test cases displayed in figure \ref{lst:roci2test}, which adds an observation point (a call to  \texttt{toString()}) on instances of \texttt{StringWriter}, which are modified by \texttt{toJson()}.

The next case study is in listing \ref{lst:roci4}: two variants of the method \texttt{decode()} in the \texttt{Base64} class of the Apache Commons Codec library. 
The original program has a \texttt{switch-case} statement in which case 1 execute a break. An original comment by the programmers indicates that it is probably impossible. 
The test case in listing \ref{lst:ampli_test4} amplifies one of the original test case with a mutation on the String value in the \texttt{encodedInt3} variable (the original String has an additional `$\backslash$' character, removed by the ``remove character'' transformation). The amplification on the observation points adds multiple observations points. 
The single observation point shown in the listing is the one that detects computational diversity: it calls the static \texttt{decodeInteger()} method which returns 1 on the original program and 0 on the variant.  In addition to validating our approach, this example anecdotally answers the question of the programmer, case 1 is possible, it can be triggered from the API.

These three case examples are meant to give the reader a better idea of how DSpot was able to detect the variants. 
We discuss how augmented test cases reveal this diversity (both with amplified inputs and observation points).
We illustrate three categories of code variations that maintain the expected functionality as specified in the test suite, but still induce diversity (different checks on input, different formatting, different handling of special cases).

\begin{framed}
The diversity that we observe originates from   areas of the code that are characterized by their  flexibility (caching, checking, formatting,  etc.). These areas are very close to the concept of  \emph{forgiving region} proposed by Martin Rinard \cite{Rinard12}.
\end{framed}

\subsection{Threats to Validity}

\begin{lstlisting}[caption={Two variants of \texttt{decode} in commons.codec},label=lst:roci4,float,language=java,numbers=left]
// Original program
void decode(final byte[] in, int inPos, final int inAvail, final Context context) {
  switch (context.modulus) {
    case 0 : // impossible, as excluded above
    case 1 : // 6 bits - ignore entirely
        // not currently tested; perhaps it is impossible?
             break;
}

// variant
void decode(final byte[] in, int inPos, final int inAvail, final Context context) {
  switch (context.modulus) {
    case 0 : // impossible, as excluded above
    case 1 : 
}
\end{lstlisting}

\begin{lstlisting}[caption={Amplified test case that reveals the computational diversity between variants of listing \ref{lst:roci4}},label=lst:ampli_test4,float,language=java,numbers=left]
@Test
void testCodeInteger3_literalMutation222() {
  String encodedInt3 = 
    "FKIhdgaG5LGKiEtF1vHy4f3y700zaD6QwDS3IrNVGzNp2" 
    + "rY+1LFWTK6D44AyiC1n8uWz1itkYMZF0aKDK0Yjg==";
  Logger.logAssertArgument(Base64.decodeInteger(encodedInt3.getBytes(Charsets.UTF_8)));
}}
\end{lstlisting}

\toolname is able to effectively detect NVP-diversity using test suite amplification.
Our experimental results are subject to the following threats.

First, this experiment is highly computational, a bug in our evaluation code may invalidate our findings. However, since we have manually checked a sample of cases (the case studies of Section \ref{sec:natural:randomness} and Section \ref{sec:nature-computational-diversity}) we have a high confidence in our results. Our implementation is publicly available  \footnote{\url{http://diversify-project.github.io/test-suite-amplification.html}}.

Second, we have forged the computationally diverse program variants. Eventually, as shown on Table \ref{tab:diversity-measures}, our technique \toolname is able to detect them all. The reason is that we had a bias towards our technique when forging those variants. This is true for all self-made evaluations.
This threat on the results of the comparative evaluation against TDR is mitigated by the analytical comparison of the two approaches. Both the input space and the output space of TDR (respectively an integer tuple and a returned value) are simpler and less powerful than  our amplification technique.

Third, our experiments consider one programming language (Java) and \nbprograms different application domains. To further assess the external validity of our results, new experiments are required on different technologies and more application domains.

\section{Related work}
\label{sec:related}

The work presented is related to two main areas: the identification of similarities or diversity in source code and the automatic augmentation of test suites.

\textbf{Computational diversity} The recent work by Carzaniga et al. \cite{Carzaniga15} has a similar intent as ours: automatically identifying dissimilarities in the execution of code fragments that are functionally similar. They use random test cases generated by Evosuite to get execution traces and log the internals of the execution (executed code and the read/write operations on data). 
The main difference with our work is that they assess computational diversity and with random testing instead of test amplification. 

Koopman and DeVale~\cite{koopman1999} aim at quantifying the diversity among a set of implementations of the POSIX operating system, with respect to their responses to exceptional conditions. 
Diversity quantification in this context is used to detect which versions of POSIX provide the most different failure profiles and should thus be assembled to ensure fault tolerance. 
Their approach relies on Ballista to generate millions of input data and the outputs are analyzed to quantify the difference. 
This is an example of diversity assessment with intensive fuzz testing and observation points on crashing states.

Many other works look for semantic equivalence or diversity through static or dynamic analysis. 
Gabel and Su \cite{gabel10} investigate the level of granularity at which diversity emerges in source code. Their main finding is that, for sequences up to 40 tokens, there is a lot of redundancy. Beyond this (of course fuzzy) threshold, the diversity and uniqueness of source code appears. 
Higo and Kusumoto~\cite{higo14} investigate the interplay between structural similarity, vocabulary similarity and method name similarity, to assess functional similarity between methods in Java programs. 
They show that many contextual factors influence the ability of these similarity measures to spot functional similarity (e.g., the number of methods that share the same name, or the fact that two methods with similar structure are in the same class or not).
Jiang and Su \cite{jiang09}  extract code fragments of a given length and randomly generate input data for these snippets. Then, they identify the snippets that produce the same output values (which are considered functionally equivalent, w.r.t the set of random test inputs). They show that this method identifies redundancies that static clone detection does not find. 
Kawaguchi and colleagues~\cite{kawaguchi2010conditional} focus on the introduction of changes that break the interface behavior.  
They also use a notion of partial equivalence, where ``two versions of a program need only be semantically equivalent under a subset of all inputs''.
Gao and colleagues~\cite{gao2008} propose a graph-based analysis to identify semantic differences in binary code. This work is based on the extraction of call graphs and control flow graphs of both variants and on comparisons between these graphs in order to spot the semantic variations. 
Person and colleagues~\cite{person2008} developed differential symbolic execution, which can be used to detect and characterize behavioral differences between program versions.

\textbf{Test suite amplification} In the area of test suite amplification, the work by Yoo and Harman \cite{yoo2012} is the most closely related to our approach, and we used as the baseline for computational diversity assessment. 
They amplify test suites only with transformations on integer values, while we also transform boolean and String literals, as well as statements  test cases. 
Yoo and Harman also have two additional parameters for test case transformation: the interaction level that determines the number of simultaneous transformation on the same test case, and the search radius that bounds their search process when trying to improve the effectiveness of augmented test suites. 
Their original intent is to increase the input space coverage to improve test effectiveness. They do not handle the oracle problem in that work.

Xie~\cite{xie06} augments test suites for Java program with new test cases that are automatically generated and he automatically generates assertions for these new test cases, which can check for regression errors. 
Harder et al.~\cite{Harder03} propose to retrieve \emph{operational abstractions}, i.e., invariant properties that hold for a set of test cases. 
These abstractions are then used to compute operational differences, which detects diversity among a set of test cases (and not among a set of implementations as in our case). 
While the authors mention that operational differencing can be used to augment a test suite, the generation of new test cases is out of this work's scope. 
Zhang and Elbaum~\cite{zhang2012} focus on test cases that verify error handling code. Instead of directly amplifying the test cases as we propose, they transform the program under test: they instrument the target program by mocking the external resource that can throw exceptions, which allow them to amplify the space of exceptional behaviors exposed to the test cases. 
Pezze et al.~\cite{pezze2013} use the information provided in unit test cases about object creation and initialization to build composite test cases that focus on interactions between classes. Their main result is that the new test cases find faults that could not be revealed by the unit test cases that provided the basic material for the synthesis of composite test cases. 
Xu et al.~\cite{xu2011hybrid} refer to ``test suite augmentation'' as the following process: in case a program P evolves into P', identify the parts of P' that need new test cases and generate these tests. They combine concolic and search-based test generation to automate this process. This hybrid approach is more effective than each technique separately, but with increased costs. 
Dallmeier et al.~\cite{dallmeier2010} automatically amplify test suites by adding and removing method calls in JUnit test cases. Their objective is to produce test cases that cover a wider set of execution states than the original test suite in order to improve the quality of models reverse engineered from the code.

\section{Conclusion}
\label{sec:conclusion}

In this paper, we have presented \toolname, a novel technique for detecting one kind of computational diversity between a pair of programs.
This technique is based on test suite amplification: the automatic transformation of the original test suite. 
\toolname uses two kinds of transformations, for respectively exploring new  points in the program's input  space and exploring new observation points on the execution state. after execution with the given input points. 

Our evaluation on large open-source projects shows that test suites amplified by \toolname are capable of assessing computational diversity and that  our amplification strategy is better than the closest related work, a technique called TDR by Yoo and Harman \cite{yoo2012}. We have also presented a deep qualitative analysis of our empirical findings. Behind the performance of \toolname, our results shed an original light on the specified and unspecified parts of real-world test suites and the natural randomness of computation.

This opens avenues for future work. There is a relation between the natural randomness of computation and the so-called flaky tests (those tests that occasionally fail). To use, the assertions of the flaky tests are at the border of the natural undeterministic parts of the execution: sometimes they hit it, sometimes they don't. With such a view, we imagine an approach that characterizes this limit and proposes an automatic refactoring of the flaky tests so that they get farther from the limit of the natural randomness and enter again into the good, old and reassuring world of determinism.

\section{Acknowledgements}
This work is partially supported by the EU FP7-ICT-2011-9 No. 600654 DIVERSIFY project.

\balance  
\bibliographystyle{abbrv}
\bibliography{references}

\end{document}